\documentclass[11pt,letterpaper]{article}
\usepackage[T1]{fontenc}
\usepackage{mlmodern}
\providecommand{\mathbfit}[1]{\boldsymbol{#1}}
\usepackage{cite}
\usepackage{amsmath,amssymb,graphicx,booktabs}
\usepackage[hidelinks]{hyperref}
\newtheorem{theorem}{Theorem}
\newtheorem{proposition}[theorem]{Proposition}
\newtheorem{remark}[theorem]{Remark}
\newtheorem{lemma}[theorem]{Lemma}
\newtheorem{example}[theorem]{Example}

\newtheorem{corollary}[theorem]{Corollary}
\newcommand{\myqed}{\rule{1ex}{1ex}}

\title{Asymptotically good construction of EAQECC with controlled amount of
  entanglement}
\date{3 November 2026}

\usepackage{authblk}
\author[1]{Ryutaroh Matsumoto}
\author[2]{Ferruh \"Ozbudak}
\affil[1]{Institute of Science Tokyo\thanks{ryutaroh@ict.e.titech.ac.jp}}
\affil[2]{Sabanci University, FENS, Istanbul\thanks{ferruh.ozbudak@sabanciuniv.edu}}

\begin{document}
\maketitle
\begin{abstract}
  We show a construction of asymptotically good linear codes
  with controlled dimensions of Euclidean hulls.
  By using it and Steane's enlargement construction for EAQECCs
  (entanglement-assisted quantum error-correcting codes),
  we propose a construction of asymptotically good EAQECCs
  with a controllable amount of entanglement,
  in contrast to the previous asymptotic constructions of EAQECCs
  whose amounts of entanglement were fixed to the maximal,
  while a smaller amount of entanglement is generally preferred in
  quantum error correction.
  We also determine the asymptotic parameters of the proposed EAQECCs.
  For every square $q > 9$, the information rate is at least
  $1+\theta-2/(\sqrt{q}-1)-(2q+1)\delta/(q+1)$ on a large interval,
  where $\delta$ is the relative distance and $\theta$ is the entanglement
  ratio, and for every square $q \geq 361$ this exceeds
  the entanglement-assisted quantum Gilbert--Varshamov bound
  on an explicit interval.
  By field expansion and concatenation, we also obtain asymptotically good
  EAQECCs with a controllable amount of entanglement over every finite field,
  including the binary field.
\end{abstract}

\medskip


\medskip


\medskip

\section{Introduction}\label{s1}
Explicit constructions of asymptotically good linear codes
have been an important topic in coding theory.
In 1982, Tsfasman, Vl\v{a}du\c{t} and Zink \cite{tsfasman82}
showed the existence of asymptotically good algebraic geometry codes
exceeding the Gilbert--Varshamov bound over large finite fields.
Their bound on asymptotically good linear codes
is now called the TVZ bound.
For quantum error correction,
also by using algebraic geometry codes,
several asymptotically good constructions were proposed
\cite{ashikhmin01,chen01b,matsumoto02}.
The asymptotic constructions of quantum codes
were improved by concatenation \cite{hamada08,fan23}
over certain ranges of parameters.

It is now well-known that quantum error-correcting codes
can be constructed from self-orthogonal (or dual-containing)
linear spaces over finite fields
\cite{gottesman96,calderbank97,calderbank98,ashikhmin01,ketkar06}.
In general, the self-orthogonal constraint is difficult to be satisfied.
To overcome this difficulty,
the entanglement-assisted quantum error-correcting codes
(EAQECCs) were proposed \cite{brun06} for qubits and
later extended to qudits \cite{luo17}.
An EAQECC can be constructed from any linear code with no
self-orthogonality constraint.
In an EAQECC, its encoder and decoder share $c$ pairs of
a maximally entangled state.
The relation between the amount $c$ of entanglement and
its underlying linear code had been somewhat unclear,
but in \cite{hulleaqecc18,galindo19},
the relation $c = \dim C - \dim C \cap C^\perp$
was established (see also \cite{hsieh07,wilde08}),
where $C$ is a classical linear code defining an EAQECC
and $\perp$ denotes taking the dual code of $C$ with respect to an inner product.
The hull (with respect to a chosen inner product) of $C$
is $C \cap C^\perp$, and its dimension $\dim C \cap C^\perp$ is called the hull dimension.
When the hull dimension is zero,
the corresponding EAQECC is said to have maximal entanglement (ME),
while the maximal hull dimension $\dim C = \dim C \cap C^\perp$
corresponds to a quantum error-correcting code with no entanglement.
It was also revealed that the hull dimension can be decreased \cite{andersen24,chen23,luo24} and increased under certain conditions \cite{andersen24}
  by replacing the original codes with their equivalent codes
while preserving code parameters.
We emphasize that decreasing and increasing the hull dimension
are very different problems.
For $q \geq 4$, the hull dimension of a linear code over $\mathbf{F}_q$
can be decreased to any smaller value by passing to an equivalent code
\cite{chen23,luo24}, and this is the reason why maximal entanglement is
always available.
On the other hand, increasing the hull dimension is possible only
under restrictive conditions \cite{andersen24},
and it is impossible for many classes of codes:
for example, for $q=2$ and $q=3$ the hull dimension is invariant
under code equivalence,
and in general a linear code is equivalent to a self-orthogonal code
only in very special cases
(see \cite{chen23} for the hull-variation problem).
Consequently, the results on increasing the hull dimension do not give
EAQECCs with a smaller amount of entanglement from the asymptotically good
constructions with maximal entanglement \cite{pereira21,fan25},
and in this paper the control of the hull dimension is built into
the construction itself.
Starting from an
  asymptotically good self-orthogonal code $C \subseteq C^\perp$
  \cite{stichtenoth06}, one can find
  an equivalent code $(C')^\perp$ with the same parameters as $C^\perp$
  and an arbitrarily chosen amount of entanglement
  $\dim C' - \dim C' \cap (C')^\perp$.
  One can also use $(C')^\perp$ in place of $C'$.
  As we will show in Section \ref{s4},
  when the relative distance is larger than the entanglement ratio,
  the parameters of an EAQECC given by $C'$ or $(C')^\perp$
  are strictly worse than those of our construction.
  The superiority of our construction comes from
  the use of Steane's enlargement construction for EAQECCs \cite{galindo23},
  which requires some conditions on linear codes that are not ensured
  by merely finding an equivalent code $C'$ as above.

The existence of asymptotically good EAQECCs was shown in
\cite{qian15,hulleaqecc18}.
Recently, an asymptotically good construction of EAQECCs by
algebraic geometry codes was proposed
\cite{pereira21}, and later improved by concatenation \cite{fan25}.
Both constructions \cite{pereira21,fan25} had maximal entanglement
and could not adjust their amounts of entanglement to values other than the maximal.
Since the generation and
preservation of shared entanglement are difficult within the current technology \cite{wehner18},
shared entanglement between an encoder and a decoder is a costly and precious
resource in EAQECCs. Therefore, it is desirable to have an asymptotically good construction
with a controllable amount of entanglement.

It seems difficult to construct asymptotically good linear codes
attaining the TVZ bound with  control of their hull dimensions
while also allowing for
self-orthogonal or dual-containing codes.
In \cite{matsumoto02}, TVZ-attaining asymptotically good self-orthogonal
linear codes
were explicitly
shown and they were used for construction of asymptotically good binary quantum codes. The construction \cite{matsumoto02} did not work when the characteristic of
finite fields is odd.
Later Stichtenoth showed TVZ-attaining asymptotically good self-orthogonal
linear codes over every finite field with square number of elements \cite{stichtenoth06}.
We will use algebraic function fields and their differentials in \cite{stichtenoth06},
and modify constructions of linear codes in \cite{stichtenoth06}.
Our modified linear codes will attain the  TVZ bound asymptotically
and will also enable control over their hull dimensions
with respect to the Euclidean inner product,
as shown in this paper.

In the previous constructions of asymptotically good quantum codes
\cite{ashikhmin01,chen01b,matsumoto02}, Steane's enlargement construction
\cite{steane99,cohen99,ling10} tended to provide better asymptotic
parameters of quantum codes
than the CSS construction \cite{steane96,calderbank96}.
We will propose asymptotically good EAQECCs with control over the amount
of entanglement by the TVZ-attaining asymptotically good linear codes
and Steane's enlargement construction for EAQECCs \cite{galindo23},
and evaluate their asymptotic parameters.

The main idea is as follows.
If the code field $\mathbf{F}_q$ is large enough and $q$ is a square,
then the self-dual codes of Stichtenoth \cite{stichtenoth06}
attaining the TVZ bound can be modified so that their hull dimensions
are prescribed, and the resulting EAQECCs have, at the same time,
a controlled amount of entanglement and
the best known asymptotic relative distance and information rate
on a large interval.
We note that \cite{stichtenoth06} shows that the TVZ bound is attained by
self-dual codes, that is, in the classical setting the best known
asymptotic parameters are obtained without any entanglement,
combined with \cite{ling10,steane99}.
Here we also use entanglement and its amount is prescribed,
whereas the previous asymptotic constructions \cite{pereira21,fan25}
work only for maximal entanglement.
More precisely, for every square $q > 9$ and for every $\theta$ and $\delta$
with
$0 \leq \frac{q+1}{q}\theta \leq \delta \leq \frac{1+\theta}{2} - \frac{1}{\sqrt{q}-1}$
and $\delta > 0$,
we obtain asymptotically good $q$-ary EAQECCs with entanglement ratio $\theta$,
relative distance at least $\delta$ and information rate at least
\[
 1+\theta-\frac{2}{\sqrt{q}-1}-\frac{2q+1}{q+1}\delta
\]
(Theorem \ref{thm3}).
For every square $q \geq 361$, this exceeds the entanglement-assisted
quantum Gilbert--Varshamov bound \cite{galindo19}
when $11/30 \leq \delta \leq 11/25$ and $0 \leq \theta \leq q\delta/(q+1)$,
and for every square $q \geq 1024$,
when $1/6 \leq \delta \leq 9/20$ and $0 \leq \theta \leq q\delta/(q+1)$
(Corollary \ref{cor4}).
To the best of our knowledge, these are the best known asymptotic
parameters of EAQECCs with a prescribed amount of entanglement.
Our results are good also over small finite fields:
by field expansion (Theorems \ref{thm2} and \ref{thm:small})
and by concatenation (Theorem \ref{thm5}),
we obtain asymptotically good EAQECCs with a controlled amount of
entanglement over every finite field $\mathbf{F}_{q_0}$,
for example over the binary field (Examples \ref{ex1} and \ref{ex2}).

An extended abstract of this paper was presented at WCC 2026
\cite{matsumoto26}.
Compared with \cite{matsumoto26}, the places $P_0$ and $P_\infty$
in Section \ref{s3} are now two reserved rational places over $Q_1$,
and Lemma \ref{lem1}, the proof of Theorem \ref{thm2},
Remark \ref{rem:order} and Sections \ref{s41}--\ref{s44} are new.

This paper is organized as follows:
In Section \ref{s2}, we review previous research results that will be used later.
In Section \ref{s3}, we construct asymptotically good linear codes
with control over their Euclidean hull dimension. After that, by using them
and Steane's enlargement construction, we propose an asymptotically good
construction of EAQECCs with controlled amount of entanglement.
In Section  \ref{s4}, we clarify asymptotic parameters of EAQECCs in Section \ref{s3},
and give their improvements by field expansion and concatenation.

\section{Preliminaries}\label{s2}
\subsection{Review of Steane's enlargement for EAQECCs}\label{sec1}
In this paper, the inner product is the Euclidean
unless otherwise specified,
and
$C^\perp$ denotes the dual code of a linear code $C$ with
respect to the Euclidean inner product,
while the symplectic inner product will be used in Section \ref{sec2} below.
Let $q$ be a prime power, and by $\mathbf{F}_q$ we denote the
finite field with $q$ elements.

The following is a review of \cite[Section 2.1]{galindo23}.
Let $C$ be an $[n, k]_q$ linear code and
$c = \dim C - \dim (C \cap C^\perp)$, and assume $n + c - 2k > 0$.
One can construct an
\begin{equation}
  [[n, n+c-2k, w_H(C^\perp \setminus C); c]]_q \label{eq0}
\end{equation}
EAQECC from $C$, where $w_H$ denotes the minimum Hamming weight
of nonzero vectors in a set.

Choose $C_t \subseteq C \cap C^\perp$ such that $\dim C_t \geq 2$ and
decompose $C = C_r \oplus C_t$ as a direct sum of linear spaces.
Let $B_r$ be a generator matrix of $C_r$ and
$B_t$ be a generator matrix of $C_t$ in \cite[Theorem 7]{galindo23}.
Then by \cite[Theorem 7]{galindo23},
we can have an EAQECC with parameter
\begin{equation}
  [[n, n+c-2 \dim C_r - \dim C_t, d'; c]]_q, \label{eq1}
\end{equation}
where
\begin{equation}
  d' \geq \min\left\{  w_H(C^\perp), \left\lceil\frac{1+q}{q}w_H(C_r^\perp)\right\rceil\right\}. \label{eq2}
\end{equation}
Observe that the number of information symbols is increased by
$\dim C_t$.

The number of information symbols $n+c-2 \dim C_r - \dim C_t$
is maximized by taking $C_t = C \cap C^\perp$.
In such a case we have $\dim C_r=c$, $\dim C_t=k-c$, and
(\ref{eq1}) becomes
\begin{equation}
  [[n, n-k, d'; c]]_q. \label{eq3}
\end{equation}

\subsection{Converting codes over $\mathbf{F}_{q^m}$ to those over $\mathbf{F}_q$}\label{sec2}
In the constructions of asymptotically good binary quantum codes
without entanglement \cite{ashikhmin01,chen01b,matsumoto02},
linear codes over $\mathbf{F}_{q^m}$ are converted to
those over $\mathbf{F}_{q}$ while preserving the Euclidean
inner product. We will also use that technique and review it in this section.
In the construction of  the EAQECC in (\ref{eq0})
from the linear code $C \subset \mathbf{F}_q^n$,
one considers $C \times C \subset \mathbf{F}_q^{2n}$,
then uses $C \times C$ for defining a group of Pauli matrices
on $\mathbf{C}^{q^n}$.

For vectors $\mathbfit{x} = (x_1$, \ldots, $x_{2n})$ and
$\mathbfit{y} = (y_1$, \ldots, $y_{2n}) \in \mathbf{F}_q^{2n}$,
in Section \ref{sec2} we consider their symplectic inner product
\begin{equation}
\langle \mathbfit{x}, \mathbfit{y} \rangle_s
= \sum_{i=1}^n (x_iy_{n+i} - x_{n+i}y_i), \label{eq:s}
\end{equation}
and the symplectic weight
\[
w_s(\mathbfit{x}) = |\{ i \in \{1, \ldots, n\} : (x_i, x_{n+i}) \neq (0,0) \}|.
\]
For an $\mathbf{F}_q$-linear code $C_s \subset \mathbf{F}_q^{2n}$,
by $C_s^{\perp s}$ we denote the orthogonal space of the code $C_s$ in $\mathbf{F}_q^{2n}$
with respect to the symplectic inner product (\ref{eq:s}).
Then $C_s$ defines an
$[[n, n - \dim C_s + \frac{1}{2}(\dim C_s - \dim C_s \cap C_s^{\perp s}),
    w_s(C_s^{\perp s}\setminus C_s); \frac{1}{2}(\dim C_s - \dim C_s \cap C_s^{\perp s})]]_q$
EAQECC, where $w_s(C_s^{\perp s}\setminus C_s)$
denotes the minimum symplectic weight
$\min\{w_s(\mathbfit{x})$ : $\mathbfit{x} \in C_s^{\perp s}\setminus C_s \}$.
\cite{galindo19}.

Let $C_{s,m} \subset \mathbf{F}_{q^m}^{2n}$ be
an $\mathbf{F}_{q^m}$-linear space. By using \cite{ashikhmin00}
or \cite{galindo19},
from $C_{s,m}$ one can construct an
$[[nm, m(n - \dim C_{s,m} + \frac{1}{2}(\dim C_{s,m} - \dim C_{s,m} \cap C_{s,m}^{\perp s})),
    d; \frac{m}{2}(\dim C_{s,m} - \dim C_{s,m} \cap C_{s,m}^{\perp s})]]_q$ EAQECC
with $d \geq w_s(C_{s,m}^{\perp s}\setminus C_{s,m})$,
where the dimensions are counted as $\mathbf{F}_{q^m}$-linear spaces.
The result in Section \ref{sec2} will be useful in Section \ref{s4}
when it is combined as a kind of concatenation for
the construction of $q$-ary quantum codes with a small or non-square $q$.

The conversion of linear codes over $\mathbf{F}_{q^m}$ to those over
$\mathbf{F}_q$ preserving the Euclidean inner product is given by
the following lemma, which will be used in Section \ref{s4}.
Lemma \ref{lem1} is classical.
A self-dual basis of $\mathbf{F}_q$ over $\mathbf{F}_{q_0}$, that is, a basis
with $\beta = 1$ in Lemma \ref{lem1}, exists if and only if $q_0$ is even or
$m$ is odd \cite{seroussi80,imamura83,jungnickel90}
(see also \cite[Section 2.3]{lidl97}); in the remaining case the lemma follows
from the classification of nondegenerate symmetric bilinear forms over finite
fields of odd characteristic \cite[Lemma 1]{jungnickel90} and the surjectivity
of the norm \cite[Theorem 2.28]{lidl97}.
We include a short proof for completeness.

\begin{lemma}\label{lem1}
Let $q_0$ be a prime power, $m \geq 1$ and $q = q_0^m$.
There exist $\beta \in \mathbf{F}_q^*$ and a basis
$e_1$, \ldots, $e_m$ of $\mathbf{F}_q$ over $\mathbf{F}_{q_0}$
such that
\[
 \mathrm{Tr}_{\mathbf{F}_q/\mathbf{F}_{q_0}}(\beta e_i e_j) = \delta_{ij}
 \quad \mbox{for } 1 \leq i, j \leq m.
\]
Let $\Phi : \mathbf{F}_q^n \rightarrow \mathbf{F}_{q_0}^{mn}$ be the
$\mathbf{F}_{q_0}$-linear map replacing each coordinate
$x = \sum_{i=1}^m x_i e_i$ ($x_i \in \mathbf{F}_{q_0}$) by $(x_1, \ldots, x_m)$.
Then for every $\mathbf{F}_q$-linear code $U \subseteq \mathbf{F}_q^n$
we have
\[
 \Phi(U)^\perp = \Phi(U^\perp), \quad
 \dim_{\mathbf{F}_{q_0}} \Phi(U) = m \dim_{\mathbf{F}_q} U, \quad
 w_H(\Phi(U)) \geq w_H(U).
\]
\end{lemma}
\textbf{Proof.}
If $q_0$ is even, then there is a self-dual basis of
$\mathbf{F}_q$ over $\mathbf{F}_{q_0}$ \cite{seroussi80}
(see also \cite[Section 2.3]{lidl97}), and we take $\beta = 1$.
Let $q_0$ be odd. For $\beta \in \mathbf{F}_q^*$, the symmetric bilinear form
$(x, y) \mapsto \mathrm{Tr}_{\mathbf{F}_q/\mathbf{F}_{q_0}}(\beta x y)$
on $\mathbf{F}_q$ is nondegenerate \cite[Theorem 2.24]{lidl97},
hence by \cite[Theorem 6.21]{lidl97} there is a basis $u_1$, \ldots, $u_m$
with $\mathrm{Tr}_{\mathbf{F}_q/\mathbf{F}_{q_0}}(\beta u_i u_j) = c_i \delta_{ij}$,
$c_i \in \mathbf{F}_{q_0}^*$.
Up to a nonzero square factor, $c_1 \cdots c_m$ is the determinant of the
Gram matrix of this form in any basis, which is
$N_{\mathbf{F}_q/\mathbf{F}_{q_0}}(\beta)$ times the determinant for $\beta=1$.
Since the norm is surjective onto $\mathbf{F}_{q_0}^*$ \cite[Theorem 2.28]{lidl97},
we choose $\beta$ such that $c_1 \cdots c_m$ is a square,
so that the number of nonsquares among the $c_i$ is even.
If $c_i$ is a square, we replace $u_i$ by $u_i/\sqrt{c_i}$.
The nonsquare $c_i$ are paired; for such a pair we may assume
$c_i = c_j = \nu$ after multiplying $u_j$ by a suitable element of
$\mathbf{F}_{q_0}^*$, and we replace $(u_i, u_j)$ by
$(a u_i + b u_j, -b u_i + a u_j)$, where $a, b \in \mathbf{F}_{q_0}$
satisfy $a^2 + b^2 = \nu^{-1}$; such $a$, $b$ exist since every element of
$\mathbf{F}_{q_0}$ is a sum of two squares.
This gives the required basis.
For $\mathbfit{x}, \mathbfit{y} \in \mathbf{F}_q^n$ we have
$\Phi(\mathbfit{x}) \cdot \Phi(\mathbfit{y})
 = \mathrm{Tr}_{\mathbf{F}_q/\mathbf{F}_{q_0}}(\beta\, \mathbfit{x} \cdot \mathbfit{y})$,
hence $\Phi(U^\perp) \subseteq \Phi(U)^\perp$, and the two spaces have
the same $\mathbf{F}_{q_0}$-dimension $m(n - \dim U)$ since $\Phi$ is injective.
The last assertion is clear. \myqed

\section{Asymptotically good linear codes with control over their hull dimensions}\label{sec3}\label{s3}
\subsection{Towers of algebraic function fields suitable for our construction}
We need a tower of algebraic function fields on which algebraic geometry codes
allow us to control their hull dimension.
One can use \cite{matsumoto02} for even characteristic,
but it does not work for odd characteristic.
We will use \cite{stichtenoth06} as it works for every positive characteristic.

We will review \cite{stichtenoth06}.
Hereafter $q$ is assumed to be a square in addition to being a prime power.
Let $F_0 = \mathbf{F}_q(x_0)$ be the rational function field.
For $i\geq 0$, let $F_{i+1}= F_i(x_{i+1})$ with
\[
x_{i+1}^{\sqrt{q}} + x_{i+1} = \frac{x_i^{\sqrt{q}}}{x_i^{\sqrt{q}-1}+1}.
\]
This is the tower of Garcia and Stichtenoth \cite{garcia96}.
We define
\[
w = x_0^{\sqrt{q}} + x_0, \quad z=w^{\sqrt{q}-1}.
\]
We now define a tower $E_0 \subset E_1 \subset \cdots $ of algebraic function fields.
Let $E_0 = \mathbf{F}_q(z)$,
and for $i\geq 1$, $E_i$ is defined as the Galois closure of $F_{i-1}/E_0$, that is,
the Galois closure of
$\mathbf{F}_q(x_0$, $x_1$, \ldots, $x_{i-1})/ \mathbf{F}_q(z)$.

Hereafter we fix an integer $i\geq2$,
and divisors and valuations are always considered in $E_i/\mathbf{F}_q$
unless otherwise stated.
We will use the following properties of $E_i/\mathbf{F}_q$.
\begin{proposition}\cite{stichtenoth06}\label{prop1}
  For $\alpha\in \mathbf{F}_q$, by $Q_\alpha$ we denote the zero of $z-\alpha$
  in $E_0=\mathbf{F}_q(z)$, and by $Q_\infty$ the pole of $z$ in $E_0$.
  Define a differential $\eta$ by
  \[
  \eta = \frac{dw}{1-z}.
  \]
  Finally by $n_i+2$ we denote the extension degree of $E_i/E_0$.
  We have the following properties:
  \begin{enumerate}
  \item The extension degree $n_i+2$ of $E_i/E_0$ is $(\sqrt{q}-1)\sqrt{q^i}p^{t(i)}$,
    where $t(i)$ is a nonnegative integer and $p$ is the characteristic of $\mathbf{F}_q$.
  \item The place $Q_1$ of $E_0$ splits completely in $E_i/E_0$.
    We reserve two of the places above $Q_1$ as $P_\infty$ and $P_0$,
    and denote the remaining places as $P_1$, \ldots, $P_{n_i}$.
  \item The reserved places $P_\infty$ and $P_0$ are distinct places of degree one
    in $E_i$ lying over $Q_1$; neither is used for evaluation.
  \item Let $g_i$ be the genus of $E_i$. Then $n_i/g_i$ converges to
    $\sqrt{q}-1$ as $i \rightarrow \infty$.
  \item There exists a divisor $G_0\geq 0$ of $E_i$, whose support
    contains no place lying over $Q_1$, such that
    $(\eta) = 2G_0 - (P_1 + \cdots +P_{n_i})-P_0-P_\infty$,
    which implies $\deg G_0 = g_i+n_i/2$.
  \item For every place $P$ of $E_i$ lying over $Q_1$, in particular for
    $P = P_j$ with $1 \leq j \leq n_i$, we have
    $\mathrm{res}_{P}(\eta) \in \mathbf{F}_{\sqrt{q}}\setminus \{0\}$.
  \item Let $W_i = A^{(i)} + B^{(i)}$, where $A^{(i)}$ and $B^{(i)}$ are the
    positive divisors of $E_i$ in \cite[Theorem~1.7]{stichtenoth06},
    supported on the zeros and on the poles of $w$, respectively.
    Then the support of $W_i$ contains no place lying over $Q_1$,
    $\deg W_i = \frac{n_i+2}{\sqrt{q}-1} + 1 - g_i$,
    and hence $\deg W_i/n_i \rightarrow 0$ as $i \rightarrow \infty$.
  \end{enumerate}
\end{proposition}
The two reserved places are taken from the completely split fiber, so
the evaluation length is reduced by only two. The genus limit and item 7 follow
from \cite[Theorem~1.7]{stichtenoth06}: there we have
$(w) = e_0^{(i)} A^{(i)} - e_\infty^{(i)} B^{(i)}$ with positive integers
$e_0^{(i)}$, $e_\infty^{(i)}$, so that the support of $W_i$ consists of
zeros and poles of $w$, where $z = w^{\sqrt{q}-1} \neq 1$, and
$g_i = [E_i : \mathbf{F}_q(w)] + 1 - \deg W_i$ with
$[E_i : \mathbf{F}_q(w)] = (n_i+2)/(\sqrt{q}-1)$.
The differential and residues are those of \cite[Proposition~4.4]{stichtenoth06}.

For a divisor $G$ of $E_i$ whose support does not contain any of the $P_j$,
we define
\[
C(G) = \{ ( f(P_1), \ldots, f(P_{n_i})) : f \in \mathcal{L}(G) \}.
\]
Let $\mathbfit{b}=(b_1$, \ldots, $b_{n_i}) = (\mathrm{res}_{P_1}(\eta)$, \ldots,
$\mathrm{res}_{P_{n_i}}(\eta)) \in (\mathbf{F}_{\sqrt{q}}\setminus \{0\})^{n_i}$,
and fix  $\mathbfit{a} =(a_1$, \ldots, $a_{n_i}) \in \mathbf{F}_q^{n_i}$
such that the component-wise product (Schur product)
of $\mathbfit{a}$ with itself is $\mathbfit{b}$, that is,
\begin{equation}
  a_j^2 = b_j \label{eq:ab}
\end{equation}
for $j=1$, \ldots, $n_i$.
Note that there are $2^{n_i}$ choices of $\mathbfit{a}$ for odd $q$,
while $\mathbfit{a}$ is unique for even $q$.

The following proposition was essentially given in \cite{stichtenoth06}.
\begin{proposition}\label{prop2}
  For any divisor $G$ whose support does not contain any of the $P_j$,
  we have $(\mathbfit{a} \cdot C(G_0+G))^\perp = \mathbfit{a} \cdot C(G_0-G-P_0-P_\infty)$,
  where for a set $C \subset \mathbf{F}_q^{n_i}$, $\mathbfit{a} \cdot C$
  denotes the set of component-wise products of $\mathbfit{a}$ and
  $\mathbfit{c} \in C$.
\end{proposition}
\textbf{Proof.}
This is essentially shown in \cite{stichtenoth06},
but here a proof is given for completeness.
By \cite[Corollary 2.7]{stichtenoth88} we have
\begin{eqnarray*}
  C(G_0+G)^\perp &=& \mathbfit{b} \cdot C(P_1+\cdots+P_{n_i}+(\eta)-G-G_0) \\
  &=& \mathbfit{b} \cdot C(G_0-G-P_0-P_\infty).
\end{eqnarray*}
The relation (\ref{eq:ab}) between $\mathbfit{a}$ and $\mathbfit{b}$
implies $C(G_0+G)^\perp = \mathbfit{a} \cdot \mathbfit{a} \cdot C(G_0-G-P_0-P_\infty)$,
which completes the proof.
\rule{1ex}{1ex}

\subsection{Asymptotically good EAQECCs with controlled amount of
  entanglement}
Let $\ell_0$ be a nonnegative integer and $\ell_\infty$ a positive
integer.
We will use $\mathbfit{a}\cdot C(G_0 - \ell_\infty P_\infty + \ell_0 P_0)$
as $C$ in Section \ref{sec1}.
Then by Proposition \ref{prop2}, we have
\begin{eqnarray*}
  C^\perp &=& \mathbfit{a}\cdot C(G_0+(\ell_\infty-1)P_\infty-(\ell_0+1)P_0),\\
  C \cap C^\perp &=& \mathbfit{a}\cdot C(G_0-\ell_\infty P_\infty-(\ell_0+1)P_0).
\end{eqnarray*}
For the second equality, note that under (\ref{eq100}) and (\ref{eq4}) below
the divisor $G_0+(\ell_\infty-1)P_\infty+\ell_0P_0$ has degree at most $n_i-3$,
so that evaluation is injective on its Riemann--Roch space.
Let
\begin{equation}
  \ell_r \geq 2+\ell_\infty. \label{eq100}
\end{equation}
If we use
$\mathbfit{a}\cdot C(G_0 - \ell_r P_\infty + \ell_0 P_0)$ as $C_r$ in Section \ref{sec1},
then we have $C_r^\perp = \mathbfit a\cdot C(G_0+(\ell_r-1)P_\infty-(\ell_0+1)P_0)$,
and we use an $(\ell_r - \ell_\infty)$-dimensional space contained
in $\{ \mathbf{0} \} \cup (\mathbfit{a}\cdot C(G_0-\ell_\infty P_\infty-(\ell_0+1)P_0) \setminus \mathbfit{a}\cdot C(G_0 - \ell_r P_\infty - (\ell_0+1)P_0))$ as $C_t$.

\begin{proposition}\label{prop3}
  Let $f_j \in \mathcal{L}(G_0 - (\ell_r-j) P_\infty - (\ell_0+1)P_0) \setminus \mathcal{L}(G_0 - (\ell_r-j+1) P_\infty - (\ell_0+1)P_0)$ for $j=1$, \ldots, $\ell_r - \ell_\infty$, 
 and let $C'_t$ be a linear space spanned by
$(f_j(P_1)$, \ldots, $f_j(P_{n_i}))$ for $j=1$, \ldots, $\ell_r - \ell_\infty$,
and $C_t = \mathbfit{a} \cdot C'_t$.
If
  \begin{eqnarray}
    2g_i -1 & \leq & \deg (G_0 - \ell_r P_\infty - (\ell_0+1)P_0),\label{eq4}\\
    n_i - 1 & \geq & \deg(G_0 - \ell_\infty P_\infty + \ell_0 P_0), \label{eq5}
  \end{eqnarray}
  then we have
  \begin{eqnarray}
    C_t \cap C_r & = & \{\mathbf{0}\},\label{eq6}\\
    C_t + C_r  & =  & C. \label{eq70}
  \end{eqnarray}
\end{proposition}
\textbf{Proof.}
By (\ref{eq5}), evaluation of functions in $\mathcal{L}(G_0 - \ell_\infty P_\infty + \ell_0 P_0)$ is a linear isomorphism to $C$.
By (\ref{eq4}), for all $1 \leq j \leq \ell_r - \ell_\infty$,
we have $\mathcal{L}(G_0 - (\ell_r-j) P_\infty - (\ell_0+1)P_0) \neq \mathcal{L}(G_0 - (\ell_r-j+1) P_\infty - (\ell_0+1)P_0)$.
So we can define  $f_j \in \mathcal{L}(G_0 - (\ell_r-j) P_\infty - (\ell_0+1)P_0) \setminus \mathcal{L}(G_0 - (\ell_r-j+1)P_\infty - (\ell_0+1)P_0)$ as above for $j=1$, \ldots, $\ell_r - \ell_\infty$,
and we also have  $v_{P_\infty}(f_j) = \ell_r -j-v_{P_\infty}(G_0)$.
By the strict triangle inequality of the valuation at $P_\infty$ \cite{bn:stichtenoth},
$f_1$, \ldots, $f_{\ell_r - \ell_\infty}$ are linearly independent modulo
$\mathcal{L}(G_0 - \ell_r P_\infty + \ell_0 P_0)$.
Then $C_t \cap C_r = \{\mathbf{0}\}$.
By (\ref{eq4}) and (\ref{eq5}) we see $\dim C - \dim C_r = \ell_r - \ell_\infty$,
which is equal to $\dim C_t$, which in turn implies $C=C_t+C_r$.
\rule{1ex}{1ex}

\begin{theorem}\label{thm1}
  Assume $\ell_0 \geq 0$, $\ell_\infty \geq 1$, (\ref{eq100}), (\ref{eq4}) and that
  \begin{eqnarray}
    \deg(G_0 - \ell_\infty P_\infty + \ell_0 P_0) &\leq& n_i - 1, \label{eq7}\\
    \deg(G_0 - \ell_\infty P_\infty - (\ell_0+1)P_0) &\geq & 2g_i - 1, \label{eq8}\\
    \deg(G_0 - \ell_r P_\infty + \ell_0 P_0) &\geq & 2g_i - 1. \label{eq9}
  \end{eqnarray}
  By using $C$, $C_r$, $C_t$ defined above, and the method in Section \ref{sec1},
  one can construct an EAQECC with parameters $[[n_i, k_i, d_i; c_i]]_q$ with
  \begin{eqnarray}
    c_i     &=& \dim C(G_0 - \ell_\infty P_\infty + \ell_0 P_0)\nonumber\\
    &&{}-\dim C(G_0-\ell_\infty P_\infty-(\ell_0+1)P_0)\nonumber\\
    &=& 2\ell_0+1,\nonumber\\
     k_i &=& c_i + n_i - 2\dim C_r - \dim C_t \nonumber\\
    &=& c_i + n_i -2(g_i+n_i/2-\ell_r+\ell_0-g_i+1) - (\ell_r - \ell_\infty) \nonumber\\
    &=& \ell_r+\ell_\infty-1,\label{eq202}\\
    d_i  &\geq& \min\left\{ w_H(C^\perp),
    \left\lceil\frac{q+1}{q}w_H(C_r^\perp)\right\rceil \right\}\nonumber\\
    &=& \min\biggl\{ w_H(C(G_0+(\ell_\infty-1)P_\infty-(\ell_0+1)P_0)),\nonumber\\
    &&\quad\left\lceil\frac{q+1}{q}w_H(C(G_0+(\ell_r-1)P_\infty-(\ell_0+1)P_0))\right\rceil \biggr\}\nonumber\\
    &\geq& \min \biggl\{ n_i - (g_i+n_i/2+\ell_\infty-\ell_0-2),\nonumber\\
    && \mbox{ }
    \biggl\lceil\frac{q+1}{q}[ n_i - (g_i+n_i/2+\ell_r-\ell_0-2)]\biggr\rceil\biggr\}\nonumber\\
    &=& \min \biggl\{ 2+n_i/2 - g_i -\ell_\infty + \ell_0,\nonumber\\
    &&\quad\left\lceil\frac{q+1}{q}(2+n_i/2 - g_i - \ell_r + \ell_0)\right\rceil\biggr\}.\label{eq11}
  \end{eqnarray}
  By considering $\deg G_0 = n_i/2+g_i$ and (\ref{eq100})--(\ref{eq5}),
  the condition
  \begin{equation}
    \ell_r + \ell_0 \leq \frac{n_i}{2} - g_i \label{eq10}
  \end{equation}
  is sufficient for (\ref{eq4}) and (\ref{eq7})--(\ref{eq9}) to hold.
  \rule{1ex}{1ex}
\end{theorem}

\begin{remark}\label{rem:odd}
The amount $c_i = 2\ell_0 + 1$ of entanglement in Theorem \ref{thm1}
is a positive odd integer.
In particular, the EAQECCs of Theorem \ref{thm1} always use entanglement,
and the entanglement ratio $\theta = 0$ in Section \ref{s4}
means $c_i/n_i \rightarrow 0$, not $c_i = 0$.
\end{remark}

\section{Asymptotic parameters of the proposed construction}\label{s4}
If $q_0$ is not a square, we first construct the codes over a square
extension field of $\mathbf{F}_{q_0}$ and then convert them to
$\mathbf{F}_{q_0}$-linear codes by Lemma \ref{lem1}, as in the constructions
of asymptotically good binary quantum codes
(see also \cite{ashikhmin01,chen01b,matsumoto02}).

Suppose that we will construct a $q_0$-ary asymptotically good EAQECC,
for a prime power $q_0$.
For some positive integer $m$, let $q=q_0^m$ be a square.
Then, construct $q$-ary linear codes $C$, $C_r$, $C_t$ as in the last section,
convert them to $q_0$-ary linear codes $\Phi(C)$, $\Phi(C_r)$, $\Phi(C_t)$
by Lemma \ref{lem1}, and
finally apply the result in Section \ref{sec1}
to $\Phi(C)$, $\Phi(C_r)$, $\Phi(C_t)$.
Quantum codes of the following parameters can be obtained.

\begin{theorem}\label{thm2}
  If $\ell_0 \geq 0$, $\ell_\infty \geq 1$, (\ref{eq100}) and
  $\ell_r + \ell_0 \leq \frac{n_i}{2} - g_i$ hold,
  by using $C$, $C_r$, $C_t \subset \mathbf{F}_q^{n_i}$ in Section \ref{sec3}
  and the result in Sections \ref{sec1} and \ref{sec2},
  one can construct an EAQECC with parameters $[[m n_i, m k_i, d_i; m c_i]]_{q_0}$ with
  \begin{eqnarray}
    c_i&=&2 \ell_0+1,\nonumber \\
     k_i &=& \ell_r +\ell_\infty-1,\nonumber\\
     d_i  &\geq& \min \biggl\{ 2+n_i/2 - g_i -\ell_\infty + \ell_0,\nonumber\\
     && \mbox{ }\biggl\lceil\frac{q_0+1}{q_0}(2+ n_i/2 - g_i - \ell_r + \ell_0)\biggr\rceil\biggr\} \label{eq205}.
  \end{eqnarray}
\end{theorem}
\textbf{Proof.}
By Lemma \ref{lem1} we have
$\Phi(C)^\perp = \Phi(C^\perp)$,
$\Phi(C) \cap \Phi(C)^\perp = \Phi(C \cap C^\perp)$,
$\Phi(C_t) \subseteq \Phi(C) \cap \Phi(C)^\perp$,
$\Phi(C) = \Phi(C_r) \oplus \Phi(C_t)$,
and the $\mathbf{F}_{q_0}$-dimensions of these spaces are $m$ times
their $\mathbf{F}_q$-dimensions, in particular
$\dim \Phi(C_t) = m(\ell_r - \ell_\infty) \geq 2$.
Moreover $w_H(\Phi(C)^\perp) \geq w_H(C^\perp)$ and
$w_H(\Phi(C_r)^\perp) \geq w_H(C_r^\perp)$.
Applying (\ref{eq1}) and (\ref{eq2}) over $\mathbf{F}_{q_0}$
to $\Phi(C)$, $\Phi(C_r)$, $\Phi(C_t)$ and using Theorem \ref{thm1},
we obtain the assertion.
\rule{1ex}{1ex}

\begin{remark}\label{rem:order}
The order of the two steps in Theorem \ref{thm2} is important.
We first convert $C$, $C_r$, $C_t$ to $\mathbf{F}_{q_0}$-linear codes by
Lemma \ref{lem1} and then apply Steane's enlargement \cite{galindo23}
over $\mathbf{F}_{q_0}$, which gives the factor $\frac{q_0+1}{q_0}$ in (\ref{eq205}).
If one first applies \cite{galindo23} over $\mathbf{F}_q$ to $C$, $C_r$, $C_t$
and then converts the resulting $\mathbf{F}_q$-linear space by the result in
Section \ref{sec2}, one obtains only the factor $\frac{q+1}{q}$, which is smaller
for $m > 1$.
\end{remark}

Let $\delta'/m$ be a desired relative quantum distance,
$\gamma = g_i/ n_i$ be the genus relative to the number of evaluation points,
and $\theta$ be the ratio of entanglement divided by the code length $mn_i$.
For the construction of asymptotically good EAQECCs,
parameters are chosen as follows, ignoring fractional parts of real numbers.
\begin{eqnarray}
\ell_0 &=& n_i \theta / 2, \label{eq301}\\
\ell_\infty &=& n_i (\frac{1}{2} + \frac{\theta}{2}- \gamma - \delta' ),\label{eq302}\\
\ell_r &=& n_i (\frac{1}{2} + \frac{\theta}{2}- \gamma - \frac{q_0\delta'}{q_0+1} ).\label{eq303}
\end{eqnarray}
By the above choices, up to terms of order $1/n_i$,
the entanglement ratio is $\theta$ and the information rate $k_i/n_i$ is
\[
\frac{k_i}{n_i} = 1+\theta -2\gamma - \frac{2q_0 +1}{q_0 + 1}\delta'.
\]
Consider asymptotically good algebraic geometry codes
$C_i$ with parameters $[n_i, n_i - g_i-n_i \delta', \geq n_i \delta']_q$,
which attain the TVZ bound,
and with
$\dim C_i^\perp - \dim C_i^\perp \cap C_i = n_i \theta$,
which is the amount of entanglement in (\ref{eq0}) with $C_i^\perp$ in place of $C$.
The CSS construction with $C_i$ and the result in
Section \ref{sec2} (originally given in \cite{ashikhmin00}) give an EAQECC
with information rate $1+\theta -2 \gamma - 2\delta'$,
relative quantum distance $\delta'/m$ and relative consumption of
entanglement $\theta$.
We observe that the information rate of the proposed construction
is larger by $\frac{1}{q_0+1}\delta'$ than the straightforward application of
the combination of \cite{stichtenoth06}
  and \cite{andersen24,chen23,luo24},
  or the combination of \cite{pereira21} and \cite{andersen24}.
  The increase $\frac{1}{q_0+1}\delta'$ comes from the use
  of \cite{galindo23}.

We will clarify the possible ranges of $\theta$ and relative distance $\delta=\delta'/m$.
  By (\ref{eq10}), (\ref{eq301})  and (\ref{eq303}), the possible range of $\theta$ is
  $[0, q_0 \delta' / (q_0+1)]$.
  By (\ref{eq302}) and $\ell_\infty \geq 1$,
  we have $\delta' \leq \frac{1+\theta}{2} -\gamma$.
  The possible ranges of $\theta$ and $\delta$ are
  \begin{equation}
    0 \leq \frac{q_0+1}{mq_0} \theta \leq \delta \leq \frac{1+\theta}{2m} -\frac{\gamma}{m}, \label{eq401}
    \end{equation}
  which  is consistent with \cite{matsumoto02} with $\theta=0$.
  Note that in \cite{matsumoto02} we have $q_0=2$ and $q=2^{2m}$.

\subsection{Asymptotic lower bounds over square fields}\label{s41}
We now state the asymptotic parameters of Theorem \ref{thm1}
in the classical form of information rate versus relative distance.
In this section $q$ is a square prime power with $q > 9$ and we put
\[
 \gamma = \frac{1}{\sqrt{q}-1},
\]
so that $g_i/n_i \rightarrow \gamma$ by Proposition \ref{prop1} and $\gamma < 1/2$.
The ratios $c_i/n_i$ and $k_i/n_i$ below converge by construction.
Since $d_i/n_i \in [0,1]$, by passing to a subsequence we may assume that
$d_i/n_i$ also converges, which does not change the other limits.

\begin{theorem}\label{thm3}
Let $q > 9$ be a square prime power and $\gamma = 1/(\sqrt{q}-1)$.
Let $\theta$ and $\delta$ be real numbers with
\[
 0 \leq \theta < 1 - 2\gamma, \qquad
 0 < \delta \leq \frac{1+\theta}{2} - \gamma.
\]
Then there exists a sequence of $[[n_i, k_i, d_i; c_i]]_q$ EAQECCs
with $n_i \rightarrow \infty$ such that
\[
 \lim_{i \rightarrow \infty} \frac{c_i}{n_i} = \theta, \qquad
 \lim_{i \rightarrow \infty} \frac{d_i}{n_i} \geq \delta,
\]
and
\begin{equation}\label{eq:thm3}
 \lim_{i \rightarrow \infty} \frac{k_i}{n_i} \geq
 \begin{cases}
 1 - \theta - 2\gamma, & \mbox{if } 0 < \delta \leq \theta,\\[2pt]
 1 - 2\gamma - \delta, & \mbox{if } \theta \leq \delta \leq \dfrac{q+1}{q}\theta,\\[6pt]
 1 + \theta - 2\gamma - \dfrac{2q+1}{q+1}\delta,
   & \mbox{if } \dfrac{q+1}{q}\theta \leq \delta.
 \end{cases}
\end{equation}
\end{theorem}
\textbf{Proof.}
Put $\beta = q/(q+1)$,
$A_i = \frac{1+\theta}{2} - \frac{g_i}{n_i}$ and
$H_i = \frac{1-\theta}{2} - \frac{g_i}{n_i}$,
so that $A_i - H_i = \theta$, and let
$A = \frac{1+\theta}{2} - \gamma$ and $H = \frac{1-\theta}{2} - \gamma > 0$
be their limits.
In Theorem \ref{thm1} we choose
\begin{eqnarray*}
 \ell_0 &=& \lfloor n_i \theta/2 \rfloor,\\
 \ell_r &=& \lfloor n_i \min\{H_i, A_i - \beta\delta\} \rfloor,\\
 \ell_\infty &=& \max\{1, \min\{ \lfloor n_i(A_i - \delta) \rfloor, \ell_r - 2\}\}.
\end{eqnarray*}
Since $A - \beta\delta > A - \delta \geq 0$ and $H > 0$,
we have $\ell_r \geq 3$ for all sufficiently large $i$,
and then $\ell_\infty \geq 1$ and (\ref{eq100}) holds.
We have
$\ell_r + \ell_0 \leq n_i H_i + n_i\theta/2 = n_i/2 - g_i$,
which is (\ref{eq10}).
By $\ell_0 \geq n_i\theta/2 - 1$,
$\ell_\infty \leq \max\{1, n_i(A_i - \delta)\}$ and
$\ell_r \leq n_i(A_i - \beta\delta)$,
the two terms in (\ref{eq11}) satisfy
\begin{eqnarray*}
 2 + n_i/2 - g_i - \ell_\infty + \ell_0 &\geq& n_i\delta - n_i\max\{0, \delta - A_i\},\\
 \frac{q+1}{q}(2 + n_i/2 - g_i - \ell_r + \ell_0) &\geq& n_i\delta,
\end{eqnarray*}
and since $A_i \rightarrow A \geq \delta$, we obtain
$\lim d_i/n_i \geq \delta$.
Finally $c_i/n_i = (2\ell_0+1)/n_i \rightarrow \theta$, and
\[
 \frac{k_i}{n_i} = \frac{\ell_r + \ell_\infty - 1}{n_i}
 \longrightarrow \min\{H, A - \beta\delta\} + \min\{H, A - \delta\}.
\]
If $\delta \leq \theta$, then $A - \delta \geq H$ and the limit is $2H$.
If $\theta \leq \delta \leq \frac{q+1}{q}\theta$,
then $A - \delta \leq H \leq A - \beta\delta$ and the limit is
$H + A - \delta$.
If $\frac{q+1}{q}\theta \leq \delta$, then $A - \beta\delta \leq H$
and the limit is $2A - (1+\beta)\delta$.
This proves (\ref{eq:thm3}).
\rule{1ex}{1ex}

The three expressions in (\ref{eq:thm3}) agree at $\delta = \theta$
and at $\delta = \frac{q+1}{q}\theta$,
so the right hand side of (\ref{eq:thm3}) is continuous in $\delta$.
The third case of (\ref{eq:thm3}) is the rate obtained in
(\ref{eq301})--(\ref{eq303}) with $m=1$, and the first two cases
cover the remaining part of the range of Theorem \ref{thm1}.

We now compare Theorem \ref{thm3} with the construction obtained by
starting from the asymptotically best self-orthogonal codes
\cite{stichtenoth06} or their duals and decreasing their hull dimensions
\cite{andersen24,chen23,luo24}, which was mentioned in Section \ref{s1}.
On the same tower, we now use all $n_i + 2$ places over $Q_1$ for evaluation.
For a divisor $G$ of $E_i$ whose support contains no place lying over $Q_1$, let
\[
\widetilde{C}(G) = \{ ( f(P_1), \ldots, f(P_{n_i}), f(P_0), f(P_\infty)) : f \in \mathcal{L}(G) \},
\]
and let $\widetilde{\mathbfit{a}} = (a_1, \ldots, a_{n_i}, a_0, a_\infty) \in \mathbf{F}_q^{n_i+2}$
with $a_0^2 = \mathrm{res}_{P_0}(\eta)$ and $a_\infty^2 = \mathrm{res}_{P_\infty}(\eta)$,
which exist by Proposition \ref{prop1}.
Since $(\eta) = 2G_0 - (P_1 + \cdots + P_{n_i}) - P_0 - P_\infty$,
the proof of Proposition \ref{prop2}, with $P_1 + \cdots + P_{n_i} + P_0 + P_\infty$
in place of $P_1 + \cdots + P_{n_i}$, gives
$(\widetilde{\mathbfit{a}} \cdot \widetilde{C}(G_0 + G))^\perp = \widetilde{\mathbfit{a}} \cdot \widetilde{C}(G_0 - G)$,
and for $G = 0$ this gives the self-dual codes of \cite{stichtenoth06}.
Let $W_i$ be as in Proposition \ref{prop1}.
For an integer $s_i$ with
$g_i - n_i/2 - 1 \leq s_i \deg W_i \leq n_i/2 + 1 - g_i$, let
$B_i = \widetilde{\mathbfit{a}} \cdot \widetilde{C}(G_0 + s_i W_i)$.
Then $B_i^\perp = \widetilde{\mathbfit{a}} \cdot \widetilde{C}(G_0 - s_i W_i)$,
so that $B_i$ is self-orthogonal if $s_i \leq 0$,
$B_i$ is dual-containing if $s_i \geq 0$, and $B_i$ is self-dual if $s_i = 0$.
The dimension $\kappa_i$ of $B_i$ is $n_i/2 + 1 + s_i \deg W_i$,
we have $w_H(B_i^\perp) \geq \kappa_i - g_i + 1$,
the hull dimension of $B_i$ is $\min\{\kappa_i, n_i + 2 - \kappa_i\}$,
and these codes attain the TVZ bound.
By \cite{andersen24,chen23,luo24} one can decrease the hull dimension
of $B_i$ to any smaller value, so that the amount $c_i$ of entanglement
in (\ref{eq0}) can be any integer between
$\max\{0, 2\kappa_i - n_i - 2\}$ and $\kappa_i$.
Since $\deg W_i/n_i \rightarrow 0$, for every $r$ with $\gamma < r < 1 - \gamma$
there are such $s_i$ for all sufficiently large $i$, with $\kappa_i/n_i \rightarrow r$.
Choosing $\kappa_i/n_i \rightarrow r$ and $c_i/n_i \rightarrow \theta$
with $\theta \leq r \leq \frac{1+\theta}{2}$ and $\gamma < r < 1 - \gamma$,
(\ref{eq0}) gives EAQECCs with relative distance at least $r - \gamma$
and information rate $1 + \theta - 2r$.
Hence, for a relative distance $\delta$, the best lower bound on the
information rate obtained in this way is
\begin{equation}\label{eq:css}
 1 + \theta - 2\max\{\gamma + \delta, \theta\}.
\end{equation}

\begin{proposition}\label{prop:compare}
Let $q$, $\gamma$, $\theta$ and $\delta$ be as in Theorem \ref{thm3}.
For $\theta < \delta \leq \frac{1+\theta}{2} - \gamma$, the information rate
in (\ref{eq:thm3}) is strictly larger than (\ref{eq:css}):
the difference is $\delta - \theta$ if $\theta < \delta \leq \frac{q+1}{q}\theta$,
and it is $\frac{\delta}{q+1}$ if $\frac{q+1}{q}\theta \leq \delta$.
For $0 < \delta \leq \theta$, (\ref{eq:css}) is at least the first case of
(\ref{eq:thm3}); it equals $1-\theta$ for $0 < \delta \leq \theta - \gamma$
and $1 + \theta - 2\gamma - 2\delta$ for
$\max\{0, \theta-\gamma\} \leq \delta \leq \theta$.
\end{proposition}
\textbf{Proof.}
For $\theta < \delta$ we have $\max\{\gamma+\delta, \theta\} = \gamma + \delta$,
and (\ref{eq:css}) becomes $1 + \theta - 2\gamma - 2\delta$.
Subtracting it from the second and the third cases of (\ref{eq:thm3})
gives $\delta - \theta$ and $\frac{\delta}{q+1}$, respectively.
The remaining assertions are immediate.
\rule{1ex}{1ex}

Thus, on the whole interval $\delta > \theta$, the proposed construction
is strictly better than the construction from self-orthogonal codes with
hull decrease, with the same relative distance and the same amount of
entanglement, and the improvement is not small:
it is $\frac{\delta}{q+1}$ for all $\delta \geq \frac{q+1}{q}\theta$,
in particular for all $\delta > 0$ when $\theta = 0$.
This improvement comes from Steane's enlargement \cite{galindo23},
which requires the subcodes $C_r$, $C_t$ with the estimates
in Section \ref{s3}, and these are not provided by hull decrease.

\subsection{Comparison with the entanglement-assisted quantum GV bound}\label{s42}
For $Q > 1$ and $0 < x < 1$, let
\[
 H_Q(x) = \frac{x \log (Q-1) - x\log x - (1-x)\log(1-x)}{\log Q}.
\]
In terms of the information rate and the entanglement ratio,
the asymptotic entanglement-assisted quantum Gilbert--Varshamov (EA-GV) bound
\cite[Theorems 5--6]{galindo19} is
\begin{equation}\label{eq:eagv}
 1 + \theta - 2H_{q^2}(\delta),
\end{equation}
for $0 < \delta < 1/2$ with $2H_{q^2}(\delta) \leq 1$ and $\theta \leq H_{q^2}(\delta)$.
Within the range of Theorem \ref{thm3},
the third case of (\ref{eq:thm3}) exceeds (\ref{eq:eagv})
whenever $\theta \leq \frac{q}{q+1}\delta$ and
\begin{equation}\label{eq:eagvcond}
 2H_{q^2}(\delta) > \frac{2}{\sqrt{q}-1} + \frac{2q+1}{q+1}\delta.
\end{equation}

\begin{corollary}\label{cor4}
Let $q$ be a square prime power.
In either of the following ranges, there exist sequences of $q$-ary EAQECCs
with entanglement ratio tending to $\theta$,
relative distance at least $\delta$, and information rate at least
\begin{equation}\label{eq:cor4}
 1 + \theta - \frac{2}{\sqrt{q}-1} - \frac{2q+1}{q+1}\delta
 \; > \; 1 + \theta - 2H_{q^2}(\delta):
\end{equation}
\[
 \begin{array}{c|c|c}
 q & \delta & \theta \\ \hline
 q \geq 361 & 11/30 \leq \delta \leq 11/25 & 0 \leq \theta \leq q\delta/(q+1)\\
 q \geq 1024 & 1/6 \leq \delta \leq 9/20 & 0 \leq \theta \leq q\delta/(q+1)
 \end{array}
\]
In particular, in these ranges the information rate of the proposed
EAQECCs is strictly larger than the EA-GV bound (\ref{eq:eagv}).
For positive $\theta$, these EAQECCs have a prescribed nonmaximal
amount of entanglement.
\end{corollary}
\textbf{Proof.}
Both ranges satisfy $\theta \leq \frac{q}{q+1}\delta$ and, as checked at the end
of the proof, the hypotheses of Theorem \ref{thm3};
hence the information rate is at least the left hand side of (\ref{eq:cor4}).
Put $h(x) = -x\log x - (1-x)\log(1-x)$ with natural logarithms.
Since $\log(q^2-1) = 2\log q + \log(1-q^{-2})$,
the difference of the two sides of (\ref{eq:cor4}) is
\begin{equation}\label{eq:gap}
 \frac{h(\delta)}{\log q} - \frac{2}{\sqrt{q}-1}
 + \delta\left( \frac{1}{q+1} + \frac{\log(1-q^{-2})}{\log q}\right).
\end{equation}
The last parenthesis is positive for $q \geq 3$,
because $-\log(1-q^{-2}) < \frac{1}{q^2-1}$ and $(q-1)\log q > 1$.
Moreover $h$ is increasing on $(0, 1/2)$ and
$(\sqrt{q}-1)/\log q$ is increasing for $q > 1$.
Therefore it suffices to check
\[
 h(11/30) > \frac{\log 361}{9}, \qquad
 h(1/6) > \frac{2 \log 1024}{31},
\]
and the differences are larger than $0.0028$ and $0.0033$, respectively.
Furthermore, since $h$ is increasing on $(0,1/2)$ and $\log(1-q^{-2})<0$,
in both ranges we have
\begin{eqnarray*}
 2H_{q^2}(\delta) &<& 2\delta + \frac{h(\delta)}{\log q}
 \leq \frac{22}{25} + \frac{h(11/25)}{\log 361} < 1, \\
 2H_{q^2}(\delta) &<& \frac{9}{10} + \frac{h(9/20)}{\log 1024} < 1,
\end{eqnarray*}
respectively, and $H_{q^2}(\delta) > \delta \geq \theta$
since $h(\delta) > \delta \log 2 > -\delta \log(1-q^{-2})$,
so that (\ref{eq:eagv}) is applicable.
The hypotheses of Theorem \ref{thm3} hold since
$\theta \leq q\delta/(q+1) < 1 - 2\gamma$ and
$\delta \leq 11/25 < 1/2 - 1/18 \leq (1+\theta)/2 - \gamma$ for $q \geq 361$,
and $\delta \leq 9/20 < 1/2 - 1/31$ for $q \geq 1024$.
\rule{1ex}{1ex}

For example, with $\theta = 1/10$, we obtain the following information
rates directly over the indicated fields; the third column is the EA-GV bound
(\ref{eq:eagv}), the fourth column is the construction from self-orthogonal
codes with hull decrease (\ref{eq:css}), and the last column is the proposed
construction:
\begin{center}
\footnotesize
\setlength{\tabcolsep}{4pt}
\begin{tabular}{@{}ccccc@{}}
\toprule
$q$ & $\delta$ & EA-GV (\ref{eq:eagv}) & Self-orthogonal, hull decrease (\ref{eq:css}) & Proposed (Theorem \ref{thm3})\\
\midrule
$361$ & $2/5$ & $0.185715$ & $0.188889$ & $619/3258 \simeq 0.189994$\\
$1024$ & $1/5$ & $0.627807$ & $0.635484$ & $201987/317750 \simeq 0.635679$\\
\bottomrule
\end{tabular}
\end{center}
In both rows, the proposed construction is better than both the EA-GV bound
and the construction from self-orthogonal codes with hull decrease.
To the best of our knowledge, Theorem \ref{thm3} and Corollary \ref{cor4}
give the best known asymptotic relative distance and information rate
of EAQECCs with a prescribed amount of entanglement, for every square
$q \geq 361$ and on the indicated intervals;
the amount of entanglement can be chosen arbitrarily small.

\subsection{Smaller fields}\label{s43}
Theorem \ref{thm2} gives EAQECCs over every finite field.
Its asymptotic form is as follows.

\begin{theorem}\label{thm:small}
Let $q_0$ be a prime power, $m \geq 1$ such that $q = q_0^m$ is a square
and $q > 9$, and $\gamma_m = 1/(q_0^{m/2} - 1)$.
Let $\theta$ and $\delta$ be real numbers with
\[
 0 \leq \theta < 1 - 2\gamma_m, \qquad
 0 < m\delta \leq \frac{1+\theta}{2} - \gamma_m.
\]
Then there exists a sequence of $[[N_i, K_i, D_i; E_i]]_{q_0}$ EAQECCs
with $N_i \rightarrow \infty$ such that
$\lim E_i/N_i = \theta$, $\lim D_i/N_i \geq \delta$, and
\begin{equation}\label{eq:small}
 \lim_{i \rightarrow \infty} \frac{K_i}{N_i} \geq
 \begin{cases}
 1 - \theta - 2\gamma_m, & \mbox{if } 0 < m\delta \leq \theta,\\[2pt]
 1 - 2\gamma_m - m\delta, & \mbox{if } \theta \leq m\delta \leq \dfrac{q_0+1}{q_0}\theta,\\[6pt]
 1 + \theta - 2\gamma_m - \dfrac{2q_0+1}{q_0+1}m\delta,
   & \mbox{if } \dfrac{q_0+1}{q_0}\theta \leq m\delta.
 \end{cases}
\end{equation}
\end{theorem}
\textbf{Proof.}
Apply Theorem \ref{thm2} with the choices of $\ell_0$, $\ell_\infty$, $\ell_r$
in the proof of Theorem \ref{thm3},
where $\gamma_m$, $m\delta$ and $q_0/(q_0+1)$ are used in place of
$\gamma$, $\delta$ and $\beta$.
Then $N_i = mn_i$, $K_i = mk_i$, $E_i = mc_i$, $D_i = d_i$, and
$\lim D_i/N_i \geq \delta$ since the estimates in the proof of
Theorem \ref{thm3} give $\lim d_i/n_i \geq m\delta$.
\rule{1ex}{1ex}

Here $m$ is fixed while $i \rightarrow \infty$.
Thus, for every finite field $\mathbf{F}_{q_0}$ and every
$0 \leq \theta < 1$, choosing $m$ with $\gamma_m < (1-\theta)/2$
and then a sufficiently small $\delta > 0$,
we obtain asymptotically good $q_0$-ary EAQECCs with entanglement ratio $\theta$.
The comparison in Section \ref{s41} is also valid after field expansion,
where the hull decrease is applied over $\mathbf{F}_q$ before the
field expansion by Lemma \ref{lem1}, and (\ref{eq:css}) becomes
$1 + \theta - 2\max\{\gamma_m + m\delta, \theta\}$;
in particular the proposed construction is strictly better
when $m\delta > \theta$, even for the binary field.

\begin{example}\label{ex1}
Let $q_0 = 2$, $\theta = 1/6$ and $1/40 \leq \delta \leq 1/32$.
Theorem \ref{thm:small} gives sequences of binary EAQECCs with
entanglement ratio tending to $1/6$,
relative distance at least $\delta$, and information rate at least
\[
 \begin{cases}
 \dfrac{205}{186} - \dfrac{50}{3}\delta, & \mbox{if } 1/40 \leq \delta \leq 219/8060,\\[6pt]
 \dfrac{13}{15} - 8\delta, & \mbox{if } 219/8060 \leq \delta \leq 1/32,
 \end{cases}
\]
where $m = 10$ is used in the first interval and $m = 8$ in the second.
For every admissible $m$, the bound
$1 + \theta - 2\max\{\gamma_m + m\delta, \theta\}$
is at most $31/30 - 16\delta$, with equality for $m = 8$;
indeed $m$ must be even, $m = 2$ is not admissible,
the values for $m = 4, 6$ are smaller than the value for $m = 8$,
and for even $m \geq 8$ the value decreases when $m$ is replaced by $m+2$,
since
\[
 2\left(\frac{1}{2^{m/2}-1} - \frac{1}{2^{m/2+1}-1}\right) - 4\delta
 \leq \frac{32}{465} - \frac{1}{10} < 0.
\]
Hence the displayed bound is larger by at least
$613/12090 > 0.0507$ throughout the interval,
the minimum being attained at $\delta = 219/8060$.
\end{example}

\subsection{Concatenation}\label{s44}
A further improvement over small fields is obtained by concatenation
with a short inner code.
By \cite[Theorem 1]{fan22concat}
(see \cite[Lemma 3]{fan25} for the $q_0$-ary case),
an unassisted stabilizer code $[[a, b, d_0]]_{q_0}$ used as the inner code
and an EAQECC $[[N, K, D; E]]_{q_0^b}$ used as the outer code give an
\[
 [[aN, bK, d \geq d_0 D; bE]]_{q_0}
\]
EAQECC. In particular, the amount of entanglement is still controlled
by the outer code.
By using the EAQECCs of Theorem \ref{thm:small} of this paper as the outer
codes and the concatenation rule \cite[Theorem 1]{fan22concat},
we obtain the following theorem.

\begin{theorem}\label{thm5}
Let $[[a, b, d_0]]_{q_0}$ be an unassisted stabilizer code with $b > 0$.
Put $Q = q_0^b$, and let $m \geq 1$ be such that $Q^m$ is a square
and $Q^m > 9$. For real numbers $\theta \geq 0$ and $\delta > 0$, put
\[
 \vartheta = \frac{a\theta}{b}, \qquad
 \Delta = \frac{a\delta}{d_0}, \qquad
 \gamma = \frac{1}{Q^{m/2} - 1}.
\]
If
\[
 0 \leq \vartheta < 1 - 2\gamma, \qquad
 0 < m\Delta \leq \frac{1+\vartheta}{2} - \gamma,
\]
then there exist sequences of $q_0$-ary EAQECCs with entanglement ratio
tending to $\theta$, relative distance at least $\delta$, and
information rate at least
\begin{equation}\label{eq:thm5}
 \begin{cases}
 \dfrac{b}{a} - \theta - \dfrac{2b\gamma}{a},
   & \mbox{if } m\Delta \leq \vartheta,\\[6pt]
 \dfrac{b}{a} - \dfrac{2b\gamma}{a} - \dfrac{bm}{d_0}\delta,
   & \mbox{if } \vartheta \leq m\Delta \leq \dfrac{Q+1}{Q}\vartheta,\\[6pt]
 \dfrac{b}{a} + \theta - \dfrac{2b\gamma}{a} - \dfrac{bm(2Q+1)}{d_0(Q+1)}\delta,
   & \mbox{if } \dfrac{Q+1}{Q}\vartheta \leq m\Delta.
 \end{cases}
\end{equation}
\end{theorem}
\textbf{Proof.}
Apply Theorem \ref{thm:small} over $\mathbf{F}_Q$ with entanglement ratio
$\vartheta$, relative distance $\Delta$ and extension degree $m$,
and concatenate with the inner code.
The information rate and the entanglement ratio are multiplied by $b/a$,
and the relative distance is multiplied by $d_0/a$.
Passing to a subsequence, the relative distances of the concatenated codes
converge.
\rule{1ex}{1ex}

\begin{example}\label{ex2}
Let $q_0 = 2$. In the notation of Section \ref{sec2}, let
$C_s \subset \mathbf{F}_2^{20}$ be the $\mathbf{F}_2$-linear space spanned by
$(1, \ldots, 1, 0, \ldots, 0)$ and $(0, \ldots, 0, 1, \ldots, 1)$.
Then $C_s \subseteq C_s^{\perp s}$, $C_s^{\perp s}$ consists of the vectors
$(x_1, \ldots, x_{20})$ with $\sum_{i=1}^{10} x_i = \sum_{i=11}^{20} x_i = 0$,
and $w_s(C_s^{\perp s} \setminus C_s) = 2$.
Hence $C_s$ defines an unassisted $[[10, 8, 2]]_2$ code, which we take as
the inner code.
Choose $Q = 256$, $m = 1$, $\vartheta = 1/16$ and $\Delta = 1/4$.
The third case of (\ref{eq:small}) gives outer codes over $\mathbf{F}_{256}$
with information rate at least
\[
 1 + \frac{1}{16} - \frac{2}{15} - \frac{513}{257}\cdot\frac{1}{4}
 = \frac{26531}{61680}.
\]
Hence Theorem \ref{thm5} gives binary EAQECCs with entanglement ratio
$\theta = 1/20$, relative distance at least $1/20$, and information rate at least
\[
 \frac{26531}{77100} \simeq 0.344112.
\]
For the same $\theta$ and $\delta$, the best value of (\ref{eq:small})
with $q_0 = 2$ is $37/140 \simeq 0.264286$, attained with $m = 6$.
Thus concatenation increases the information rate by
$\frac{21541}{269850} \simeq 0.079826$.
\end{example}

\section*{Declarations}
\noindent \textbf{Use of AI tools.}
The authors used ChatGPT for the optimization of the asymptotic parameters
in Section \ref{s4},
and also used ChatGPT and  Gemini for finding mathematical and
citational issues in earlier versions of this paper.
All results were checked by the authors, who take full responsibility
for the content of this paper.

\noindent \textbf{Conflict of interest.}
The authors have no conflict of interest to declare.

\noindent \textbf{Data availability.}
No data sets were generated or analyzed in this work.

\noindent \textbf{Acknowledgment.}
The first author was/is partially supported by ``Japan Society for the Promotion of Science'' (grant 23K10980) and
by access to OpenAI models through the ChatGPT for Academic Researchers program.


\begin{thebibliography}{10}

\bibitem{andersen24}
  S. E. Anderson, E. Camps-Moreno, H. H. L\'opez, G. L. Matthews, D. Ruano and I. Soprunov, Relative hulls and quantum codes.
  {\em IEEE Trans. Inform. Theory}, 70(5):3190--3201, May 2024.

\bibitem{ashikhmin00}
A.~Ashikhmin and E.~Knill.
\newblock Nonbinary quantum stabilizer codes.
\newblock {\em IEEE Trans. Inform. Theory}, 47(7):3065--3072, Nov. 2001.

\bibitem{ashikhmin01}
A.~Ashikhmin, S.~Litsyn, and M.~A. Tsfasman.
\newblock Asymptotically good quantum codes.
\newblock {\em Phys. Rev. A}, 63(3):032311, Mar. 2001.

\bibitem{brun06}
T.~Brun, I.~Devetak, and M.-H. Hsieh.
\newblock Correcting quantum errors with entanglement.
\newblock {\em Science}, 314(5798):436--439, 2006.

\bibitem{calderbank97}
A.~R. Calderbank, E.~M. Rains, P.~W. Shor, and N.~J.~A. Sloane.
\newblock Quantum error correction and orthogonal geometry.
\newblock {\em Phys. Rev. Lett.}, 78(3):405--408, Jan. 1997.

\bibitem{calderbank98}
A.~R. Calderbank, E.~M. Rains, P.~W. Shor, and N.~J.~A. Sloane.
\newblock Quantum error correction via codes over {GF(4)}.
\newblock {\em IEEE Trans. Inform. Theory}, 44(4):1369--1387, July 1998.

\bibitem{calderbank96}
A.~R. Calderbank and P.~W. Shor.
\newblock Good quantum error-correcting codes exist.
\newblock {\em Phys. Rev. A}, 54(2):1098--1105, Aug. 1996.

\bibitem{chen23}
H.~Chen.
\newblock On the hull-variation problem of equivalent linear codes.
\newblock {\em IEEE Transactions on Information Theory}, 69(5):2911--2922, May
  2023.

\bibitem{chen01b}
H.~Chen, S.~Ling, and C.~Xing.
\newblock Asymptotically good quantum codes exceeding the
  {Ashikhmin-Litsyn-Tsfasman} bound.
\newblock {\em IEEE Trans. Inform. Theory}, 47(5):2055--2058, July 2001.

\bibitem{cohen99}
G.~Cohen, S.~Encheva, and S.~Litsyn.
\newblock On binary constructions of quantum codes.
\newblock {\em IEEE Trans. Inform. Theory}, 45(7):2495--2498, Nov. 1999.

\bibitem{fan25}
J.~Fan, W.~Cheng, G.~Luo, Z.~Li, and M.~Cao.
\newblock Entanglement-assisted concatenated quantum codes: Parameters and
  asymptotic performance, 2025.
\newblock arXiv:2501.04921.

\bibitem{fan23}
J.~Fan, J.~Li, Y.~Wang, Y.~Li, M.-H. Hsieh, and J.~Du.
\newblock Partially concatenated Calderbank-Shor-Steane codes achieving the
  quantum Gilbert-Varshamov bound asymptotically.
\newblock {\em IEEE Transactions on Information Theory}, 69(1):262--272, 2023.

\bibitem{fan22concat}
J.~Fan, J.~Li, Y.~Zhou, M.-H.~Hsieh, and H.~V.~Poor.
\newblock Entanglement-assisted concatenated quantum codes.
\newblock {\em Proceedings of the National Academy of Sciences},
119(24):e2202235119, 2022.

\bibitem{galindo19}
C.~Galindo, F.~Hernando, R.~Matsumoto, and D.~Ruano.
\newblock Entanglement-assisted quantum error-correcting codes over arbitrary
  finite fields.
\newblock {\em Quantum Inf. Process}, 18:116, 2019.

\bibitem{galindo23}
C.~Galindo, F.~Hernando, and R.~Matsumoto.
\newblock Steane enlargement of entanglement-assisted quantum error-correcting
  codes.
\newblock {\em Comp. Appl. Math.}, 42:275, Aug. 2023.

\bibitem{garcia96}
A.~Garcia and H.~Stichtenoth.
\newblock On the asymptotic behaviour of some towers of function fields over
  finite fields.
\newblock {\em J. Number Theory}, 61(2):248--273, Dec. 1996.

\bibitem{gottesman96}
D.~Gottesman.
\newblock Class of quantum error-correcting codes saturating the quantum
  {H}amming bound.
\newblock {\em Phys. Rev. A}, 54(3):1862--1868, Sept. 1996.

\bibitem{hulleaqecc18}
K.~Guenda, S.~Jitman, and T.~A. Gulliver.
\newblock Constructions of good entanglement-assisted quantum error correcting
  codes.
\newblock {\em Des. Codes Cryptogr.}, 86:121--136, 2018.

\bibitem{hamada08}
M.~Hamada.
\newblock Concatenated quantum codes constructible in polynomial time:
  Efficient decoding and error correction.
\newblock {\em IEEE Transactions on Information Theory}, 54(12):5689--5704,
  2008.

\bibitem{hsieh07}
M.-H. Hsieh, I.~Devetak, and T.~A. Brun.
\newblock General entanglement-assisted quantum error-correcting codes.
\newblock {\em Phys. Rev. A}, 76(6):062313, Dec. 2007.

\bibitem{imamura83}
K.~Imamura.
\newblock On self-complementary bases of {GF($q^n$)} over {GF($q$)}.
\newblock {\em Trans. IECE Japan (Section E)}, E66:717--721, 1983.

\bibitem{jungnickel90}
D.~Jungnickel, A.~J. Menezes, and S.~A. Vanstone.
\newblock On the number of self-dual bases of {GF($q^m$)} over {GF($q$)}.
\newblock {\em Proc. Amer. Math. Soc.}, 109(1):23--29, May 1990.

\bibitem{ketkar06}
A.~Ketkar, A.~Klappenecker, S.~Kumar, and P.~K. Sarvepalli.
\newblock Nonbinary stabilizer codes over finite fields.
\newblock {\em IEEE Trans. Inform. Theory}, 52(11):4892--4914, Nov. 2006.

\bibitem{lidl97}
R.~Lidl and H.~Niederreiter.
\newblock {\em Finite Fields}, volume~20 of {\em Encyclopedia of Mathematics
  and its Applications}.
\newblock Cambridge University Press, Cambridge, 2nd edition, 1997.

\bibitem{ling10}
S.~Ling, J.~Luo, and C.~Xing.
\newblock Generalization of Steane's enlargement construction of quantum codes
  and applications.
\newblock {\em IEEE Trans. Inform. Theory}, 56(8):4080--4084, Aug. 2010.

\bibitem{luo24}
G.~Luo, M.~F. Ezerman, M.~Grassl, and S.~Ling.
\newblock Constructing quantum error-correcting codes that require a variable
  amount of entanglement.
\newblock {\em Quantum Information Processing}, 23:4, 2024.

\bibitem{luo17}
L.~Luo, Z.~Ma, Z.~Wei, and R.~Leng.
\newblock Non-binary entanglement-assisted quantum stabilizer codes.
\newblock {\em Science China Information Sciences}, 60:042501, Apr. 2017.
\newblock doi:10.1007/s11432-015-0932-y.

\bibitem{matsumoto02}
R.~Matsumoto.
\newblock Improvement of {Ashikhmin-Litsyn-Tsfasman} bound for quantum codes.
\newblock {\em IEEE Trans. Inform. Theory}, 48(7):2122--2124, July 2002.

\bibitem{matsumoto26}
R.~Matsumoto and F.~\"Ozbudak.
\newblock Asymptotically good construction of {EAQECC} with controlled amount
  of entanglement.
\newblock In {\em WCC 2026: The Fourteenth International Workshop on Coding
  and Cryptography}, Paris, June 2026.
\newblock \url{https://wcc2026.inria.fr/assets/final_versions/WCC2026_paper_44.pdf}.

\bibitem{pereira21}
F.~R.~F. Pereira, R.~Pellikaan, G.~G. La~Guardia, and F.~M. de~Assis.
\newblock Entanglement-assisted quantum codes from algebraic geometry codes.
\newblock {\em IEEE Transactions on Information Theory}, 67(11):7110--7120,
  Nov. 2021.

\bibitem{qian15}
J.~Qian and L.~Zhang.
\newblock Entanglement-assisted quantum codes from arbitrary binary linear
  codes.
\newblock {\em Des. Codes Cryptogr.}, 77(1):193--202, 2015.

\bibitem{seroussi80}
G.~Seroussi and A.~Lempel.
\newblock Factorization of symmetric matrices and trace-orthogonal bases in
  finite fields.
\newblock {\em SIAM J. Comput.}, 9(4):758--767, 1980.

\bibitem{steane96}
A.~M. Steane.
\newblock Multiple particle interference and quantum error correction.
\newblock {\em Proc. Roy. Soc. London Ser. A}, 452(1954):2551--2577, Nov. 1996.

\bibitem{steane99}
A.~M. Steane.
\newblock Enlargement of {Calderbank-Shor-Steane} quantum codes.
\newblock {\em IEEE Trans. Inform. Theory}, 45(7):2492--2495, Nov. 1999.

\bibitem{stichtenoth88}
H.~Stichtenoth.
\newblock Self-dual Goppa codes.
\newblock {\em J. Pure Appl. Algebra}, 55:199--211, Nov. 1988.


\bibitem{stichtenoth06}
H.~Stichtenoth.
\newblock Transitive and self-dual codes attaining the {Tsfasman-Vladut-Zink}
  bound.
\newblock {\em IEEE Trans. Inform. Theory}, 52(5):2218--2224, May 2006.

\bibitem{bn:stichtenoth}
H.~Stichtenoth.
\newblock {\em Algebraic Function Fields and Codes}, volume 254 of {\em
  Graduate Texts in Mathematics}.
\newblock Springer-Verlag, Berlin Heidelberg, 2nd edition, 2009.

\bibitem{tsfasman82}
M.~A. Tsfasman, S.~G. Vl\v{a}du\c{t}, and T.~Zink.
\newblock Modular curves, {Shimura} curves, and {Goppa} codes, better than
  {Varshamov-Gilbert} bound.
\newblock {\em Math. Nachr.}, 109:21--28, 1982.

\bibitem{wehner18}
       S. Wehner, D. Elkouss and R. Hanson,
       Quantum internet: A vision for the road ahead.
       \emph{Science},
       362, article ID eaam9288, 2018, DOI:10.1126/science.aam9288.

\bibitem{wilde08}
M.~M. Wilde and T.~A. Brun.
\newblock Optimal entanglement formulas for entanglement-assisted quantum
  coding.
\newblock {\em Phys. Rev. A}, 77(6):064302, June 2008.

\end{thebibliography}
\end{document}